\documentclass[prb,groupedaddress,superscriptaddress,twocolumn]{revtex4-2} 

\usepackage[T1]{fontenc}
\usepackage[utf8]{inputenc}
\usepackage{lmodern}
\usepackage{graphicx}
\usepackage{amsmath}
\usepackage{amssymb}
\usepackage{amsfonts}
\usepackage{xcolor}
\usepackage{physics}
\usepackage[colorlinks=true, linkcolor=black, citecolor=black]{hyperref}

\begin{document}
\title{Enhanced Linear Dichroism of Flattened-Edge Black Phosphorus Nanoribbons}

\author{Leandro Seixas}
  \email{leandro.seixas@mackenzie.br}
  \affiliation{MackGraphe---Graphene and Nanomaterials Research Center, Mackenzie Presbyterian University, 01302-907 S\~ao Paulo, SP, Brazil}
  \affiliation{School of Engineering, Mackenzie Presbyterian University, 01302-907 S\~ao Paulo, SP, Brazil}

\begin{abstract}
Black phosphorus is a material with an intrinsic anisotropy in electronic and optical properties due to its puckered honeycomb lattice. Optical absorption is different for incident light with linear polarization in the armchair and zigzag directions (linear dichroism). These directions are also used in the cuts of materials to create black phosphorus nanoribbons. Edges of nanoribbons usually have small reconstruction effects, with minor electronic effects. Here, we show a reconstruction of the armchair edge that introduces a new valence band, which flattens the puckered lattice and increases the linear dichroism extrinsically in the visible spectrum. This enhancement in linear dichroism is explained by the polarization selection rule, which considers the parity of the wave function to a reflection plane. The flattened-edge reconstruction originates from the inversion of chirality of the P atoms at the edges and significantly alters the entire optical absorption of the material. The flattened edges have potential applications in pseudospintronics, photodetectors and might provide new functionalities in optoelectronic and photonic devices.
\end{abstract}

\maketitle

\section{Introduction} 
The black phosphorus (BP) is an allotrope of phosphorus with an anisotropic puckered lattice stacked and bonded through van der Waals interaction. The BP crystal was first synthesized in 1914, by a phase transition of white phosphorus under high pressures\cite{bridgman1914two}. A hundred years later, a BP monolayer \cite{liu2014phosphorene} (a.k.a. phosphorene) was isolated using mechanical exfoliation methods similar to that used in graphene\cite{graphenescience,graphenermp}. This material soon drew attention because of its outstanding electronic properties, like current modulation on the order of 10$^{5}$ and field-effect mobility up to 1000~cm$^{2}$V$^{-1}$s$^{-1}$ at room temperature\cite{li2014black,koenig2014electric,koenig2016electron}. The BP is a semiconductor with widely tunable bandgap with the number of layers, going from 0.3~eV (bulk) to 2.0~eV (monolayer)\cite{carvalho2016phosphorene}. An exciting property of BP comes from its structural anisotropy. This structural anisotropy causes anisotropic electronic and optical properties, with different optical absorption for the linearly polarized light (linear dichroism)\cite{qiao2014high,xia2014rediscovering,yuan2015polarization}. Structural changes can cause variations in optical absorption and linear dichroism. This enhanced linear dichroism can also have applications creating a pseudospin polarization in high-performance photodetectors\cite{yuan2015polarization}, near-perfect polarizers, and pseudospintronic devices\cite{bp_pseudospin}.

Similarly to the graphene edges, the BP also has an armchair and zigzag edges\cite{graphene_edges,graphene_nanoribbons}. Due to the high anisotropy of the electronic and optical properties of BP, the edge effect gives rise to unique properties like new Raman modes activated by the phonon edges\cite{bp_edge_phonons}, and even magnetic orders for the zigzag edges\cite{du2015unexpected,ren2018half}. The reconstruction of the edges depends on the passivation or non-passivation with hydrogen atoms. Zigzag edges without H passivation can form reconstructions with self-passivation of P atoms\cite{lee2017atomic}. For armchair edges, passivation with H increases the bandgap compared to the bare material\cite{li2014electronic,wu2015electronic}. The BP nanoribbons (PNRs) come with armchair (APNR), and zigzag (ZPNR) edges\cite{guo2014phosphorene}. Quantum confinement effect increases the bandgap by decreasing the widths of the nanoribbons\cite{tran2014scaling}. APNR has good electrical conductivity and low thermal conductivity, serving as materials with significant potential applications in thermoelectric devices. The material can reach a figure of merits of up to ZT = 6.4 \cite{zhang2014phosphorene}. Recently, new methods for the production of narrow PNRs have emerged, creating ZPNR as narrow as 4~nm and as long as 75~$\mu$m\cite{watts2019production,liu2020unzipping}.

In this letter, we show that PNR cut in the armchair direction can have a reconstruction in which the P atoms at the edges move inverting the chirality of the tetrahedral crystal field. These edges are reconstructed by flattening and have unique electronic and optical properties.

\section{Methods} 
The atomistic simulations were performed via \textit{ab initio} calculations based on density functional theory (DFT) \cite{dft1, dft2} as implemented in the \textsc{Siesta} code\cite{siesta2020}. We used norm-conserved Troullier--Martins pseudopotentials\cite{troulliermartins}, and mesh cutoff of $400$~Ry. The atomic orbitals were based on double-$\zeta$ polarized (DZP) basis set and energy shift of $0.03$~eV for the basis confinement. The exchange-correlation functional used was in the PBEsol approximation\cite{pbesol}. We used the exchange-correlation functional with van der Waals correction parameterized by Vydrov--van Voorhis (vdW-VV) \cite{vv10} for calculations involving stacking more than one layer. We sample the Brillouin zone with the Monkhorst--Pack algorithm using $20$~k-points in the $y$-direction (armchair)\cite{monkhorstpack}. All geometries were optimized with forces smaller than $0.01$~eV/\AA. We used vacuum spacing of at least $15$~\AA\ in $x$ and $z$ directions to avoid interaction between periodic images. For the projected density of states (PDOS) calculations, we used a fine grid with $400$~k-points in the Brillouin zone, and a Gaussian smearing of $0.04$~eV. The energy barrier calculations were performed with the NEB algorithm\cite{cineb}, implemented in the \textsc{Ase} program\cite{ase}, and with forces smaller than 0.05~eV/\AA. Also, we calculated the optical properties using an optical mesh of $100$ q-points, optical broaden of $0.1$~eV, and optical scissor of $1.01$~eV for monolayers, $0.82$~eV for bilayers, and $0.81$~eV for trilayers. For the \textit{ab initio} molecular dynamics simulations, we use the NVT ensemble with a Nos\'{e} thermostat at 300 K. The dynamics were performed with 5000 steps of 0.5~fs. We also use a relaxation time of 100~fs and a Nos\'{e} mass of 100~ Ry(fs)$^2$.

\begin{figure}[!tb]
  \centering
      \includegraphics[width=0.48\textwidth]{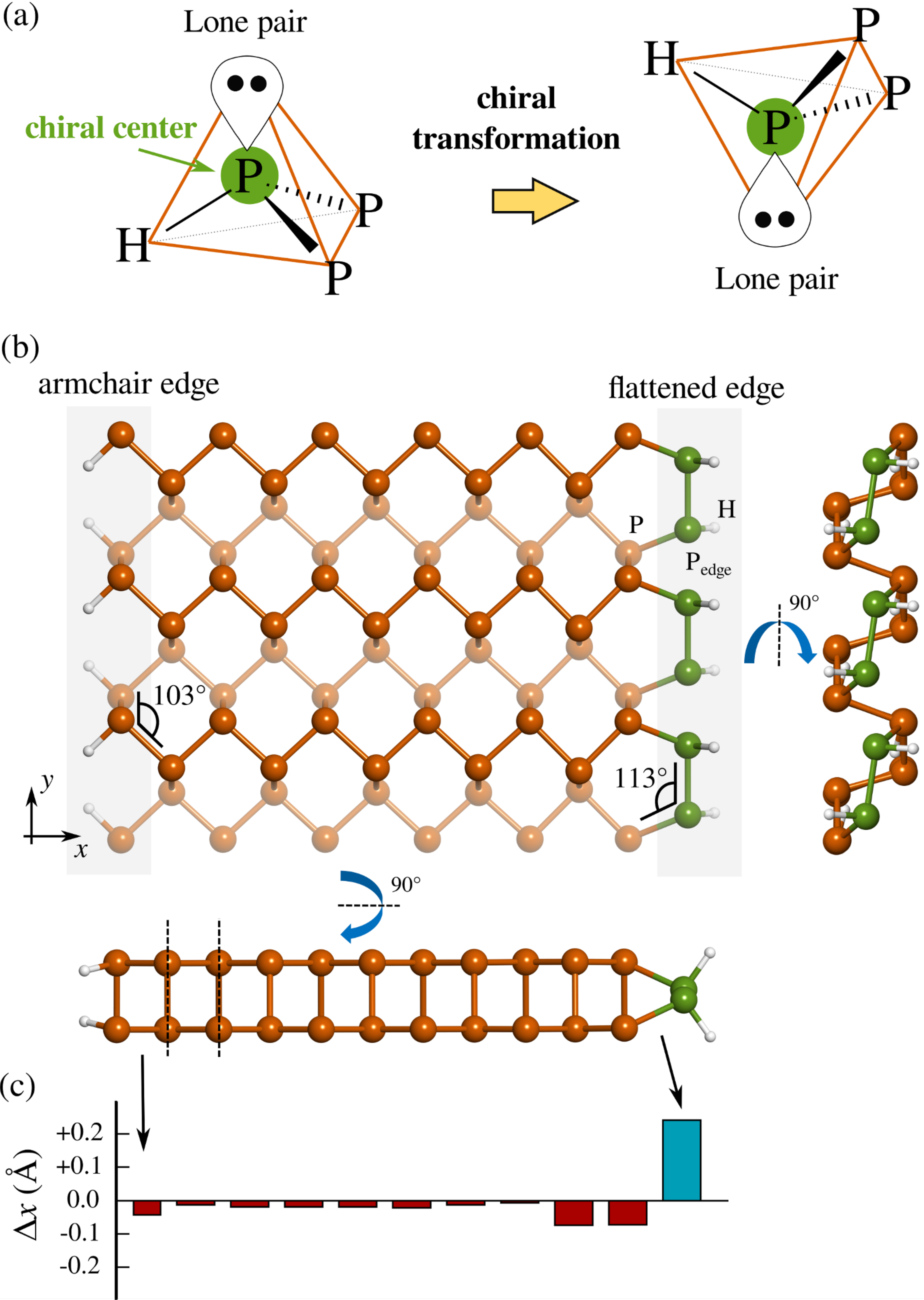}
      \caption{\textbf{Structural properties of the armchair and flattened edges}. (a) Chiral transformation of P atoms at the BP edge. (b) Ball-and-stick representation of BP with an armchair edge (left) and a flattened edge (right). (c) Variation in the $x$ coordinate with respect to the lattice constant $a = 3.33$~\AA.}
  \label{fig:Fig1}
\end{figure}

\section{Results and discussion} 
In the crystal structure of BP, the P atoms bound covalently with three other P atoms and have a lone pair, creating a tetrahedral symmetry around the atom. At the edge of the material, one of the P atoms in that tetrahedron is lost, forming a dangling bond. This dangling bond is unstable and is stabilized by passivating the edge with H atoms from the environment. The tetrahedral crystal field of the edge atoms (P$_{\rm edge}$) are shown in Fig. \ref{fig:Fig1}(a). These atoms at the border have greater mobility than those of the bulk, being able to invert the chirality by moving the H atom and the lone pair. When there is this inversion of chirality at the P$_{\rm edge}$, we can reconstruct the armchair edge with a flattening shown in Fig. \ref{fig:Fig1}(b). This reconstructed edge is called here flattened edge. After this reconstruction, the P$_{\rm edge}$ angle with the other two neighbouring P atoms is 113$^{\circ}$, greater than the 103$^{\circ}$ angle of the armchair edge, and closer to the 109.47$^{\circ}$ of the perfect tetrahedron. The variations of the $x$ coordinate for armchair and flattened edges are shown in Fig. \ref{fig:Fig1}(c). We note a small shortening for the armchair edges and an elongation of up to 0.2~\AA\ for the flattened edges.

\begin{figure*}[!tb]
  \centering
      \includegraphics[width=0.98\textwidth]{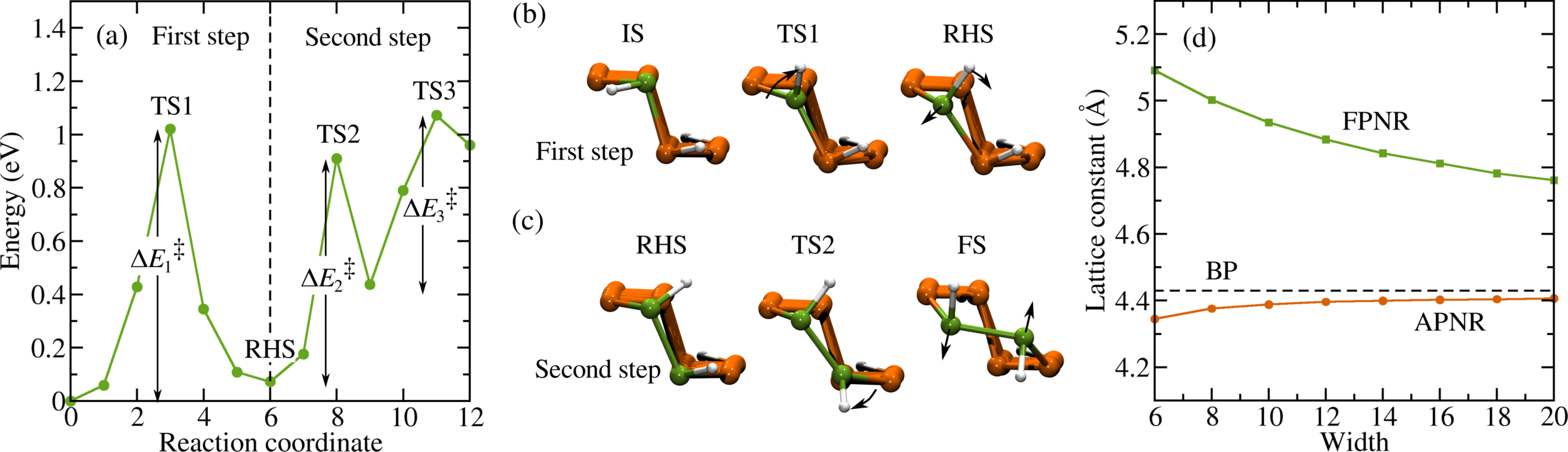}
      \caption{\textbf{Formation mechanism of the flattened edges}. (a) Energy barrier diagram for the transformation of armchair edges into flattened edges. (b) Schematic representation of reaction path for the first step, from the initial state (IS) to the rotated hydrogen state (RHS). (c) Schematic representation of reaction path for the second step, from the RHS to the final state (FS). The first and second transition states are shown: TS1 and TS2. (d) Lattice constant evolution for APNR and FPNR as function of width. The black dashed line is the lattice constant for BP monolayer.}
  \label{fig:Fig2}
\end{figure*}

\begin{figure*}[!tb]
  \centering
      \includegraphics[width=0.7\textwidth]{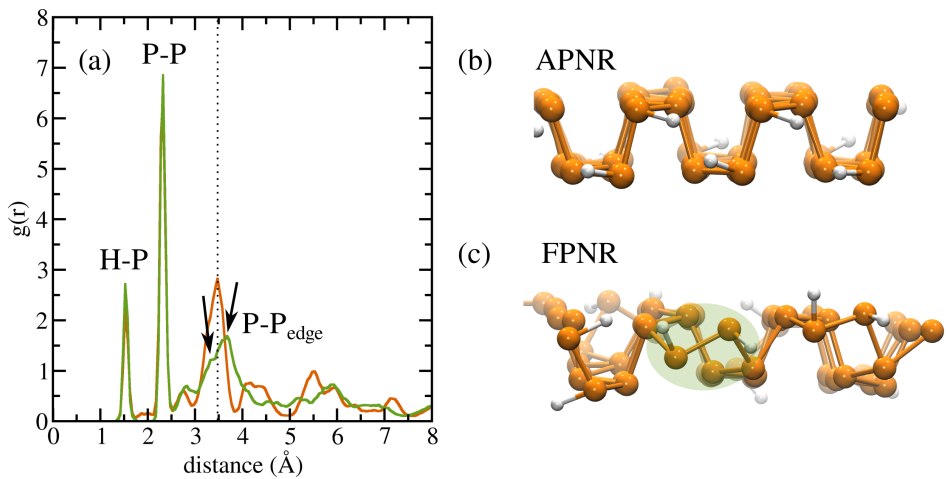}
      \caption{\textbf{\textit{Ab initio} molecular dynamics for APNR and FPNR}. (a) Radial distribution function ($g(r)$) for APNR (orange) and FPNR (green). The radial distribution was calculated with the last 1000 steps of the molecular dynamics. Reconstruction of the flattened edges is highlighted by the black arrows. Side view of BP nanoribbon geometry after 5000 steps (2.5 ps): (b) 10-APNR and (c) 10-FPNR.}
  \label{fig:Fig3}
\end{figure*}

The dynamic stability of BP nanoribbons with armchair and flattened edges is analyzed using \textit{ab initio} molecular dynamics (AIMD) at room temperature. We used three unit cells of the 10-APNR and 10-FPNR. After 5000 steps of the AIMD, we calculate the radial distribution function to check the integrity and crystallinity of the materials, as shown in Fig. \ref{fig:Fig3}(a). The first peak observed in the radial distribution is associated with the distance of the H atoms from the P atoms from the edge (H-P). The second peak observed is associated with the distances between P atoms inside the nanoribbons (P-P). Both H-P and P-P peaks are present in the radial distributions of the APNR and FPNR, also showing the crystallinity and stability of the FPNR. However, there is a crucial difference in peaks around 3.5~\AA\ (P-P$_{\textrm{edge}}$). In APNR, this peak is located around 3.47 ~\AA\ (dotted line), while in FPNR, there is a significant peak at 3.68~\AA\ and a minor bump around 3.37~\AA, highlighted by the black arrows in the Fig. \ref{fig:Fig3}(a). This duplication corresponds to the elongation and shortening of the flattened edges shown in Fig. \ref{fig:Fig1}(c). APNR has more excellent crystallinity for distances greater than 4~\AA. The FPNR has minor edge disorders that can be observed for distances greater than 4~\AA.

The side view of the APNR and FPNR geometries after 5000 steps of molecular dynamics are shown in Fig.\ref{fig:Fig3}(b,c). APNR is more crystalline, similar to optimized geometries. For the FPNR, the inversion of the chirality of the P$_{\textrm{edge}}$ that flattens the edges of the nanoribbons is still observed at room temperature. More excellent crystallinity and stability are expected for nanoribbons with lower temperatures. Pair of P flattened after the molecular dynamics are highlighted in green in Fig. \ref{fig:Fig3}(c).

The formation of the flattened edge from the armchair was studied by calculating the minimum energy path using the NEB method. In a first step, NEB simulations were performed with a single H rotating for inversion of chirality of P$_{\rm edge}$. After this movement, we find minimum local energy (RHS, rotated hydrogen state) with energy $E_{\rm RHS} = 0.07$~eV above the initial state (IS). In the transition, there is an energy barrier of $\Delta E_1^{\ddagger} = 1.02$~eV. In a second step, we start from the RHS state and move the second H from the edge. In this process, there is a barrier of $\Delta E_2^{\ddagger} = 0.84$~eV, and then another barrier of $\Delta E_{3}^{\ddagger} = 0.64$~eV. In the end, we found a final state (FS) with the flattened edges, and with a formation energy of $\Delta E_{f} = 0.96$~eV. The energies in these processes are low enough to form the flattened edges with the incidence of light. The local minimum of the final state shows that these edges can be stable even at room temperatures. The ball-and-stick representations of the initial, final, and intermediate states are shown in Fig. \ref{fig:Fig2}(b,c).

\begin{figure*}[!tb]
  \centering
      \includegraphics[width=1.0\textwidth]{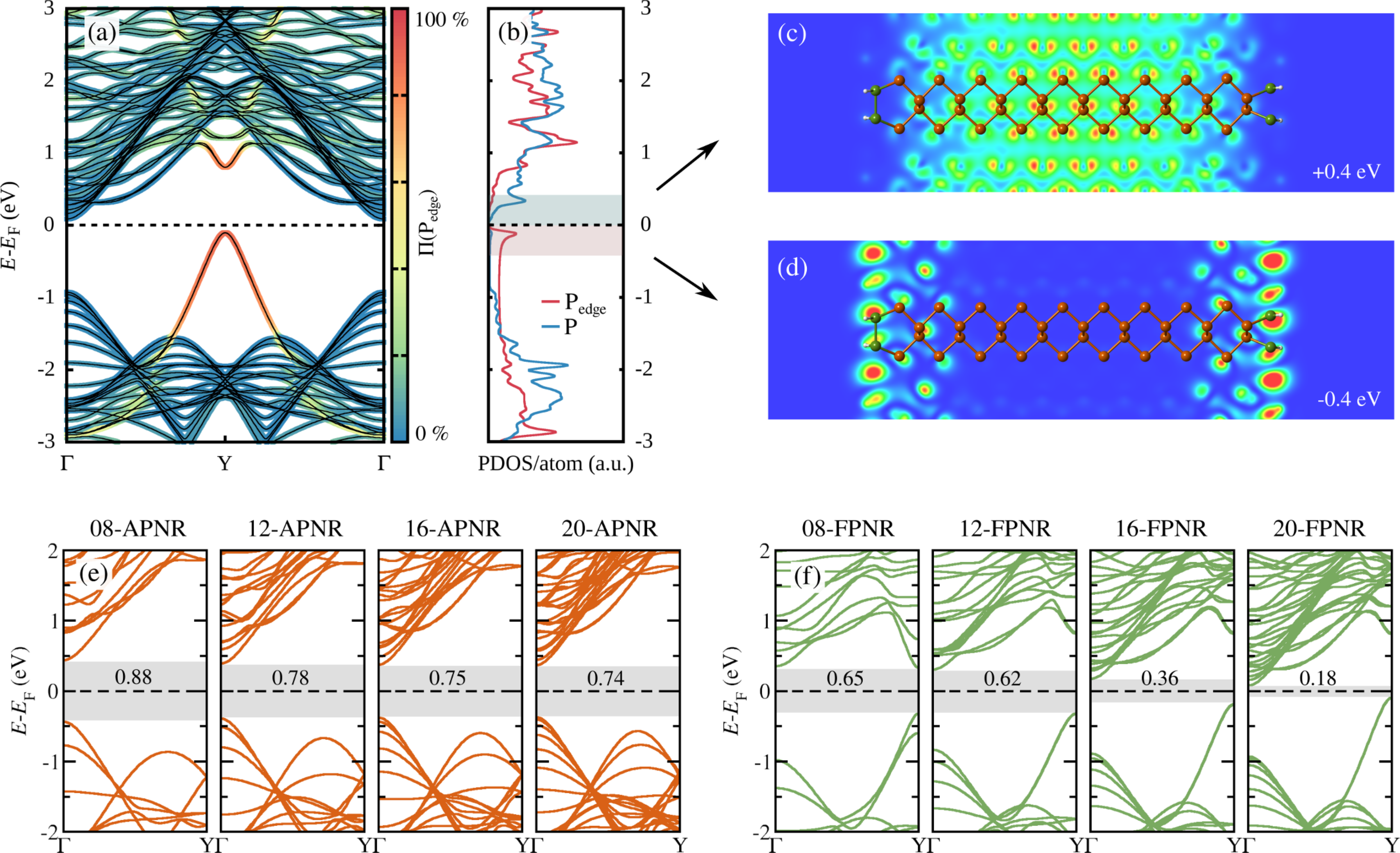}
      \caption{\textbf{Electronic properties of FPNR}. (a) Band structure with orbital decomposition on P$_{\rm edge}$ atoms. (b) Projected density of states (PDOS) per P atom. Real-space resolution of local density of states (LDOS) for: (c) electronic states between $E_F$ and $E_F+0.4$~eV (conduction band), (d) electronic states between $E_F-0.4$~eV and $E_F$ (valence band). (e) Band structures of APNR nanoribbons for four different widths. (f) Band structures of FPNR for four different widths.}
  \label{fig:Fig4}
\end{figure*}

Creating nanoribbons with armchair (APNR) and flattened (FPNR) edges, we note that the lattice constants in $y$-direction vary according to the nanoribbons' width. Similarly to the graphene nanoribbons, BP nanoribbons' widths (PNR) are labelled by the number of P dimers. A PNR with $n$ dimers wide is labeled by $n$-PNR. APNR have lattice constants slightly smaller than the BP monolayer ($b=4.42$~\AA). Meanwhile, the FPNR have lattice constants larger than BP monolayer, with values of up to 5.1~\AA\ for 6-FPNR, as shown in Fig. \ref{fig:Fig2}(d).

\begin{figure*}[!tb]
  \centering
      \includegraphics[width=0.98\textwidth]{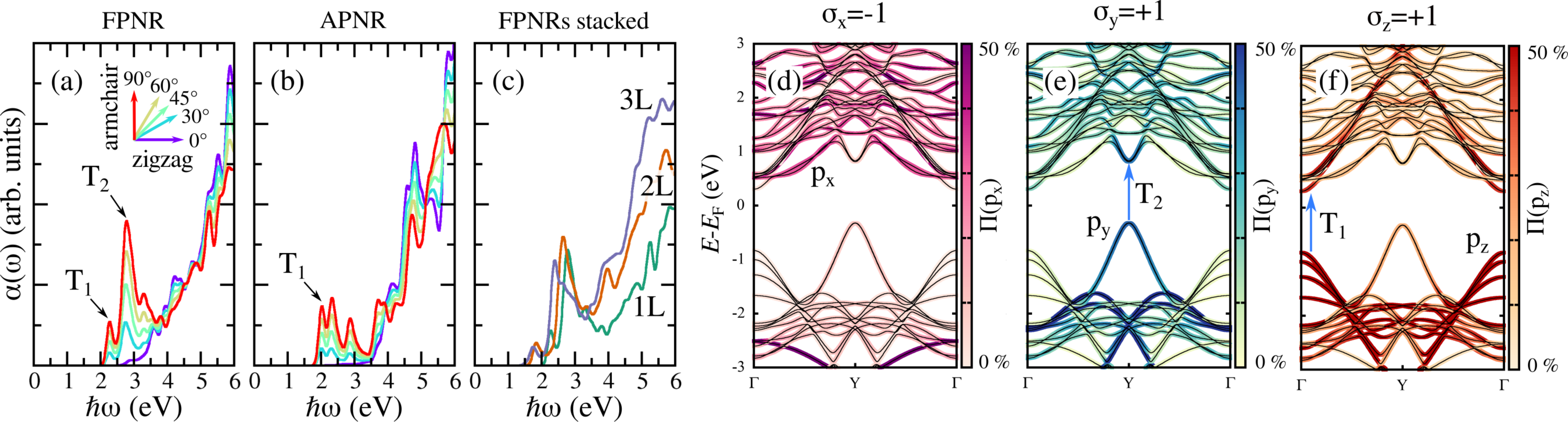}
      \caption{\textbf{Optical properties of FPNR}. Optical absorption of angle-dependent polarized light for: (a) FPNR, (b) APNR. The light polarization are rotated from 0$^{\circ}$ (zigzag) to 90$^{\circ}$ (armchair). (c) Optical absorption (unpolarized light) for one, two and three layers of FPNR stacked. Band structures with orbital projection: (d) $p_{x}$, (e) $p_{y}$, (f) $p_{z}$. The transitions $T_1$ and $T_2$ showed by the blue arrow in (e,f) are only allowed with light polarized in $y$-direction.}
  \label{fig:Fig5}
\end{figure*}

The electronic properties of FPNR differ from APNR mainly by a new parabolic valence band with maximum $Y$ point. Fig. \ref{fig:Fig4}(a) shows the 20-FPNR band structure. The indirect bandgap of this nanoribbon is $E_{\rm gap} = 0.18$~eV at the PBEsol level. For this band structure, we project the atomic orbitals that contribute to the Bloch state. For a set of atomic orbitals $\{ \ket{\phi_j}\}$, and a Bloch state $\ket{\psi_{n\vb*{k}}}$, the projection is defined by:
\begin{equation}
  \Pi(\phi) = \sum_{j}\left| \braket{\phi_j}{\psi_{n\vb*{k}}} \right|^2.
\end{equation}
Selecting the $\ket{\phi_j}$ atomic orbitals of the P$_{\rm edge}$ atoms, we calculate the quantity $\Pi(\mathrm{P_{edge}})$. We note that the valence band around $Y$ is mostly located on the flattened edges. The projected density of states (PDOS) corresponding to the band structure is shown in Fig.\ref{fig:Fig4}(b). The PDOS in Fig. \ref{fig:Fig4}(b) is normalized by the number of atoms in the projection. Selecting energy ranges in the PDOS, we can plot the local density of states (LDOS) in real space. For that, we choose two energy ranges: between the Fermi level ($E_F$) and $E_F+0.4$~eV, shown in Fig. \ref{fig:Fig4}(c), and between $E_F-0.4$~eV and $E_F$, shown in Fig. \ref{fig:Fig4}(d). The conduction band states are located in the basal plane of the FPNR, while the valence band states are located at the edges. In Fig. \ref{fig:Fig4}(c,d) higher values of LDOS values are shown in red and lower values in blue. The band structures of APNR and FPNR for some nanorribbon widths are shown in Fig. \ref{fig:Fig4}(e,f). APNRs are always semiconductors with direct bandgap at $\Gamma$ point, that vary with the width of the nanoribbon with a power law.\cite{tran2014scaling} FPNRs are indirect semiconductors for wide nanoribbons, with the conduction band minimum at the $\Gamma$ point and the valence band maximum at the $Y$ point. However, for widths less than 8-FPNR, there is a transition from indirect to direct bandgap. This transition occurs due to the interaction between border states, which result in bonding and anti-bonding states. They also vary more with nanoribbons' width, ranging from 0.65~eV (8-FPNR) to 0.18~eV (20-FPNR).

With the new states introduced in the valence band, we expect changes in the optical properties. Using the DFT with bandgap corrections calculated from the BP band gaps from the GW theory\cite{tran2014layer}, we calculated the imaginary part of the dielectric functions ($\varepsilon''(\omega)$) of the 12-FPNR and 12-APNR nanoribbons. From these dielectric functions, we calculate the optical absorption coefficient $\alpha(\omega)$. Since BP is a material with an intrinsic linear dichroism\cite{xia2014rediscovering,yuan2015polarization}, we calculate the optical absorption coefficient for various light polarization angles, from 0$^{\circ}$ (zigzag) to 90$^{\circ}$ (armchair). The absorption coefficient $\alpha(\omega)$ for 12-FPNR is shown in Fig. \ref{fig:Fig5}(a), and for 12-APNR is shown in Fig. \ref{fig:Fig5}(b). We can notice the extrinsic linear dichroism of FPNRs in the visible light spectrum, with optical absorption of the polarization in the $y$-direction (90$^{\circ}$) much greater than in the $x$-direction (0$^{\circ}$). Comparing the nanoribbons with armchair and flattened edges, we note that the absorption of polarized light in the $y$-direction is greater in FPNR than in APNR. For the optical absorption of FPNR, there is a small peak around $T_1=2.28$~eV and a second higher peak $T_2=2.78$~eV. The optical absorption around 2.8~eV is three times higher in FPNR than in APNR. Fig. \ref{fig:Fig5}(c) shows the optical absorption of one, two, and three stacked nanoribbons. We note small redshifts of the $T_2$ transition with the number of layers. The peak $T_2$ is 2.78~eV for a single layer, 2.64~eV for two layers, and 2.40~eV for three layers.

To understand the optical transitions and linear dichroism of FPNRs, we calculated the band structures with the projection of the $p_{x}$, $p_{y}$ and $p_{z}$ orbitals, as shown in Fig. \ref{fig:Fig5}(d,e,f). The linear dichroism can be explained by the parity of the wavefunction with respect to the $M_{x}$ mirror symmetry ($yz$ plane)\cite{yuan2015polarization}. The parity $\sigma_j$ for an eigenstate $\ket{\phi_j}$ is given by:
\begin{equation}
  M_{x}\ket{\phi_j} = \sigma_j \ket{\phi_j}.
\end{equation}
The mirror symmetry transform the vector $\vb{r}$ by $(x,y,z) \to (-x,y,z)$. Doing this symmetry transformation, the orbital $p_x$ has eigenvalue $\sigma_x = -1$, whereas the orbitals $p_y$ and $p_z$ have eigenvalues $\sigma_y = +1$ and $\sigma_z = +1$. For a light incident normally to the basal plane ($\vu{z}$), and polarization $\vb*{E}=E_x\vu{x}$, the mirror symmetry transform $\vb*{E}\to -\vb*{E}$  $(E_x \to -E_x)$. The matrix elements for optical transitions will only be non-zero if the valence and conduction band states have opposite parities. For example, $\sigma_x =-1 \to \sigma_y =+1$. For light with polarization $\vb*{E}=E_y\vu{y}$, the mirror symmetry transform $\vb*{E}\to \vb*{E}$ $(E_y \to E_y)$. Optical transitions are allowed only if the valence and conduction band states have the same parity. For example, $\sigma_y =+1 \to \sigma_y=+1$. This rule is called the polarization selection rule\cite{yuan2015polarization}. For nanoribbons, this mirror symmetry is broken, but there are still traces of the parity in the wavefunctions. Since the new edge states have major contributions from the $p_y$ orbitals [Fig. \ref{fig:Fig5}(e)], with parity $\sigma_y = + 1$, the optical transition between these states will only occur for polarized light in the $y$-direction, according to the expected from the optical absorption shown in Fig. \ref{fig:Fig5}(a). Although there is an intrinsic linear dichroism in BP, it can be enhanced extrinsically by the formation of flattened edges. From the analysis of the parities of the wavefunctions and linear dichroism, we see that the $T_1$ absorption peak refers to vertical transitions at the $\Gamma$-point and the $T_2$ absorption peak refers to the vertical transitions at the $Y$-point. 

\section{Conclusions} 
In conclusion, we show that an unconventional reconstruction of the BP armchair edge can be stabilized. BP nanoribbons with these new flattened edges have unique electronic and optical properties, with the emergence of a valence band with a maximum at the $Y$ point. This band converts the material from a direct bandgap to an indirect bandgap semiconductor. The optical absorption of the nanoribbons demonstrates a significant increase in the visible light spectrum, with linear dichroism enhanced by the flattened edges. This linear dichroism can have potential applications in BP bipolar pseudospintronics, high-performance photodetectors and photonic devices integrated with silicon.

\section*{Acknowledgments}
We thank Prof. Christiano J. S. de Matos and Dr. Henrique B. Ribeiro for valuable discussions on optical properties of BP and the financial support from the MackPesquisa, Conselho Nacional de Desenvolvimento Cient\'\i fico e Tecnol\'ogico (CNPq) (Grant Nos. 40825/2018-5 and 311324/2020-7). We also thank High Performance Computing Center (NACAD) at COPPE, UFRJ for providing computational facilities.


%

\end{document}